\documentclass[prl,twocolumn,10pt,titlepage]{revtex4}
\usepackage{amsmath}
\usepackage{graphicx}
\usepackage{dcolumn}
\usepackage{bm}

\input{tcilatex}

\begin{document}

\title{Two-dimensional quantum spin-$1/2$ Heisenberg model with competing
interactions}
\author{ J. Ricardo de Sousa$^{(a)}$and N. S. Branco$^{(b)}$}
\affiliation{$^{(a)}$Departamento de F\'{\i}sica, Universidade Federal do Amazonas,
69077-000, Manaus, AM - Brazil\\
$^{(b)}$Departamentio\ de F\'{\i}sica, Universidade Federal de Santa
Catarina, 88040-900, Florian\'{o}polis, SC - Brazil }

\begin{abstract}
We study the quantum spin-$1/2$ Heisenberg model in two dimensions,
interacting through a nearest-neighbor antiferromagnetic exchange ($J$) and
a ferromagnetic dipolar-like interaction ($J_{d}$), using double-time
Green's function, decoupled within the random phase approximation (RPA). We
obtain the dependence of $k_{B}T_{c}/J_{d}$ $\ $as a function of frustration
parameter $\delta $, where $T_{c}$ is the ferromagnetic (F) transition
temperature and $\delta $ is the ratio between the strengths of the exchange
and dipolar interaction (i.e., $\delta =J/J_{d}$). The transition
temperature between the F and paramagnetic phases decreases with $\delta $,
as expected, but goes to zero at a finite value of this parameter, namely $%
\delta =\delta _{c}=\pi /8$. At $T=0$ (quantum phase transition), we analyze
the critical parameter $\delta _{c}(p)$ for the general case of an exchange
interaction in the form $J_{ij}=J_{d}/r_{ij}^{p}$, where ferromagnetic and
antiferromagnetic phases are present. 
\end{abstract}

\maketitle

\section{I. Introduction}

Considerable attention during the last decade has been devoted to the
investigation of systems with long-range interactions. In particular, the
interest in films and quasi-two-dimensional systems have attracted attention
mainly due to their technological applications as, for example, in
electronics, data storage, catalysis in the case of metal-on-metal films,
biotechnology, and pharmacology in the case of molecular films. The magnetic
size unit and its thermal stability are essential points to be considered in
order to obtain a good performance in magneto-optical recording\cite{1}. The
magnetic properties of these films depend on the subtle interplay between
the long-range antiferromagnetic dipolar interaction, the short-range
rotational invariant exchange, and the magnetic surface anisotropy. The
presence of antiferromagnetic domains (AF) was observed in epitaxial thin
films by the polarization-dependent X-ray magnetic linear dichroism spectra
microscopy\cite{2}. A possible intrinsic mechanism leading to AF domains is
the competition between anisotropy and dipolar interaction\cite{3}.

In two-dimensional lattices and for spins with rotational symmetry,
long-range order does not occur at any finite temperature, for quantum and
classical models with short-range interactions\cite{4}. That is the case
when only short-range exchange interactions are present, for the XY and
Heisenberg models, for example. For all real systems there is a
long-range dipolar interaction, with the important property that it breaks
the symmetry between out-of-plane orientation of the spins and the in-plane
orientation of the spins in the ordered state. The presence of dipolar
interactions in systems with rotational symmetry (exchange) may stabilize
long-range order at finite temperatures, for both classical and quantum
models\cite{5}.

The competition between long-range antiferromagnetic dipolar and short-range
ferromagnetic exchange interactions in two-dimensional uniaxial (Ising) spin
systems is responsible for a very rich phenomenological scenario concerning
both their equilibrium statistical mechanics\cite{6,7,8,9} and
non-equilibrium dynamic properties\cite{10,1,12,13,14,15}. By means of Monte
Carlo simulations and analytical calculations of the ground state, MacIsaac
et al.\cite{6} studied the two-dimensional spin-$1/2$ Ising model with
ferromagnetic exchange and antiferromagnetic dipolar interactions, and have
shown that for $\delta \equiv J/J_{d}<0.425$, $J$ and $J_{d}$ being the
strength of exchange and dipolar interactions, respectively, the
antiferromagnetic state is stable. For $\delta >0.425$ the AF state becomes
unstable with respect to the formation of striped domain structures, i.e.,
to state configurations with spins aligned along a particular axis forming a
ferromagnetic stripe of constant width $h$, forming a superlattice in the
direction perpendicular to the stripes. Monte Carlo results \cite{8} at low
temperatures give further support to this proposal, at least for
intermediate values of $\delta $.

There are few numerical results concerning the equilibrium statistical
mechanics of quantum models with competitive exchange and dipolar
interactions in two-dimensional lattices\cite{14,15}. The two dimensional
antiferromagnetic (AF) Heisenberg model has been investigated by many
authors to explain, for example, the magnetic mechanism of high $T_{c}$
superconductivity\cite{16,17}. Antiferromagnetic fluctuations are believed
to play an important role in the superconductivity of the cuprates\cite{17},
such as La$_{2}$CuO$_{4}$, which is well described by a
quasi-two-dimensional quantum spin-1/2 Heisenberg AF model. Chandra and
Doucot\cite{18} have studied the square-lattice Heisenberg model at $T=0$
with next-nearest-exchange coupling and suggested that the AF order is
destroyed due to the competition between the nearest and next-nearest
exchange interactions. Although frustration effects are very effective to
destroy the AF order in general, they may not be effective in high $T_{c}$
materials, considering the extended nature of the holes which destroy the AF
order. A good example of quasi-two-dimensional AF materials in which the
dipolar interaction (with presence of frustration) is comparable to the
exchange coupling are the so called high-$T_{c}$ superconductors RBa$_{2}$Cu$%
_{3}$O$_{7-x}$ (where $R$ stands for rare earth)\cite{19}.

>From a theoretical point of view, the double-time Green's function (GF)
theory\cite{20} is both a convenient and an effective theoretical framework
for interpretation and forecasting of various characteristics of matter at
all temperatures. The development of approximation schemes for the GF
approach has focused on decoupling their equations of motion. This
decoupling is usually chosen for convenience or for reasons which are
essentially \textit{ad hoc}. The consistency of the basic decoupling
approximation with relevant operator identities is not always assured.
Several first-order decoupling have been proposed in the literature:
firstly, the random phase approximation (RPA) was applied to the quantum
spin-$1/2$ Heisenberg ferromagnetic\cite{20}, and extended to include a
long-range interaction $J(r)$, which depends on the distance $r$ between
spins as an inverse power law $J(r)=J/r^{p}$, where $p>d$ (dimension)\cite%
{21,22,23}. This RPA decoupling predicts the absence of magnetic order in
finite temperature and low-dimensionality ($d=1,2$) for $p>2d$, in
accordance with the generalized Mermin and Wagner theorem\cite{24}. The
quantitative results of the phase diagram in the ($T-p$) plane in the $%
d<p<2d $ region obtained by RPA \cite{23} are in accordance with quantum
Monte Carlo simulations\cite{25}.

The aim of this work is to investigate the results of the competition
between the exchange and dipolar interactions in the two-dimensional quantum
spin-1/2 Heisenberg model. In Sec. II, the model is presented and treated by
the double-time GF technique in the RPA decoupling. In Sec. III, we analyze
the phase diagram in the $T-\delta $ plane. Conclusions and remarks are
summarized in Sec. IV.

\section{II. Model and Green's Function Method}

In order to study the consequences of both quantum effects and 
frustration, we propose here a spin model to describe the destruction
of the F order, represented by the following Hamiltonian%
\begin{equation}
\mathcal{H}=J\dsum\limits_{<i,j>}\mathbf{S}_{i}\cdot \mathbf{S}%
_{j}-J_{d}\dsum\limits_{(i,j)}\frac{\mathbf{S}_{i}\cdot \mathbf{S}_{j}}{%
r_{ij}^{3}},  \label{1}
\end{equation}%
where $\mathbf{S}_{i}=(S_{i}^{x},S_{i}^{y},S_{i}^{z})$ is the spin-$1/2$
operator at site $i$. The first term, $J$, is the strength of the exchange
interaction, and the sum is over all nearest-neighbor pairs, $<i,j>$, on a
square lattice. The second term, $J_{d}=(g\mu _{B})^{2}/a^{3}$ ($g$ is the
Land\'{e} factor, $\mu _{B}$ the Bohr magneton, $a$ the lattice constant),
represents a long-ranged dipole-dipole interaction and the sum is over all
possible pairs of atoms on the square lattice. The dipolar interaction
tends to align the spins in the F system at low
temperatures ($T<T_{c}$), while the exchange interaction (antiferromagnetic, 
$J>0$) tends to destroy the long-range ferromagnetic order. Consequently the
ground state of a system determined by the dipolar interaction alone differs
from the ground state of a system determined by the exchange interaction
alone. In the absence of the dipolar interaction (i.e., $J_{d}=0$), the
isotropic Heisenberg model is regained and in two-dimensional lattices it does not
present long-range order at $T>0$ (i.e., $T_{c}=0 $). When both interactions
are present the system is inherently \textit{frustrated}.

The double-time Green's function $\left\langle \left\langle
A(t);B(0)\right\rangle \right\rangle $ is defined by\cite{20}%
\begin{equation}
\left\langle \left\langle A(t);B(0)\right\rangle \right\rangle =-i\theta
(t)\left\langle \left[ A(t),B(0)\right] \right\rangle ,  \label{2}
\end{equation}%
where $\theta (t)$ is the step function, $\left[ A,B\right] $ is the
commutator of operators $A$ and $B$, and $\left\langle \cdot \cdot \cdot
\cdot \right\rangle $ denotes an average with respect to the canonical
density matrix of the system at temperature $T$. The time-Fourier transform
of Eq. (2), $\left\langle \left\langle A;B\right\rangle \right\rangle _{E}$,
satisfies the following equation of motion%
\begin{equation}
E\left\langle \left\langle A;B\right\rangle \right\rangle _{E}=\frac{1}{2\pi 
}\left\langle \left[ A,B\right] \right\rangle +\left\langle \left\langle %
\left[ A,\mathcal{H}\right] ;B\right\rangle \right\rangle _{E},  \label{3}
\end{equation}%
and $\left\langle \left\langle \left[ A,\mathcal{H}\right] ;B\right\rangle
\right\rangle _{E}$ obeys an equation similar to Eq. (3), with a
higher-order Green function appearing on the right side. In this way, an
infinite set of coupled equations is generated. Therefore, an approximation
(decoupling) is used to obtain the Green function.

The correlation function $\left\langle BA(t)\right\rangle $ is obtained by
the spectral representation theory, which gives%
\begin{eqnarray}
\left\langle BA(t)\right\rangle  &=&\dint\limits_{-\infty }^{\infty
}J(w)e^{-iwt}dw=  \notag \\
&&\dint\limits_{-\infty }^{\infty }\frac{i\left\{ G(w+i\varepsilon
)-G(w-i\varepsilon )\right\} e^{-iwt}dw}{e^{\beta w}-1}.  \label{4}
\end{eqnarray}%
where $G(E)=\left\langle \left\langle A;B\right\rangle \right\rangle _{E}$
and $\varepsilon \rightarrow 0$. Note that Eq. (4) is the required spectral
representation for the time correlation function, where $J(w)$ is the
spectral intensity of the function $\left\langle BA(t)\right\rangle $ (in
fact, its Fourier transform \cite{20}).

With $A=S_{g}^{+}$ and $B=S_{l}^{-}$, where the spin operators are defined
by the usual commutation rules, we obtain from Eq. (3): 
\begin{eqnarray}
E\left\langle \left\langle S_{g}^{+};S_{l}^{-}\right\rangle \right\rangle
_{E} &=&\frac{m}{\pi }\delta _{gl}+2\dsum\limits_{j\neq
l}J_{jl}\{\left\langle \left\langle
S_{j}^{z}S_{g}^{+};S_{l}^{-}\right\rangle \right\rangle _{E}  \notag \\
&&-\left\langle \left\langle S_{g}^{z}S_{j}^{+};S_{l}^{-}\right\rangle
\right\rangle _{E},  \label{5}
\end{eqnarray}%
where $m=\left\langle S_{g}^{z}\right\rangle $ is the magnetization per
spin, $J_{jl}=-J$ for nearest-neighbor sites and $J_{jl}=\frac{J_{d}}{%
r_{jl}^{3}}$ for the dipolar interaction (with $\delta =J/J_{d}$).

The key problem of a first-order decoupling procedure is essentially to
express the Green's function $\left\langle \left\langle
S_{a}^{z}S_{b}^{+};S_{l}^{-}\right\rangle \right\rangle _{E}$ in terms of
lower-order Green's functions, which enables one to solve the infinite chain
of \ equations of motion in an approximate way, and that can be expressed in
the following form: 
\begin{equation}
\left\langle \left\langle S_{a}^{z}S_{b}^{+};S_{l}^{-}\right\rangle
\right\rangle _{E}\simeq m\left\langle \left\langle
S_{b}^{+};S_{l}^{-}\right\rangle \right\rangle _{E}.  \label{6}
\end{equation}

Using the decoupling (6) in Eq. (5), we obtain%
\begin{equation}
EG_{gl}(E)=\frac{m}{\pi }\delta _{gl}+2m\dsum\limits_{j\neq l}J_{jg}\left\{
G_{gl}(E)-G_{jl}(E)\right\}  \label{7}
\end{equation}%
where $G_{gl}(E)\equiv \left\langle \left\langle
S_{g}^{+};S_{l}^{-}\right\rangle \right\rangle _{E}$.

The method of calculation we use is the same as Nakano and Takahashi\cite{23}%
, and we shall not reproduce the details. Following their notation, we
obtain from Eq. (7) the Fourier transform of the Green's function $G_{K}(E)=%
\mathcal{F}\{G_{gl}(E)\}$, defined by%
\begin{equation}
G_{K}(E)=\dsum\limits_{g,l}G_{gl}(E)e^{-i\mathbf{k}.(\mathbf{r}_{g}-\mathbf{r%
}_{l})},  \label{8}
\end{equation}%
\ and, in this way, we find%
\begin{equation}
G_{K}(E)=\frac{m}{\pi (E-E_{k})},  \label{9}
\end{equation}%
where the magnon energy $E_{k}$ is%
\begin{equation}
E_{k}=2m(J_{o}-J_{k}),  \label{10}
\end{equation}%
with%
\begin{equation}
J_{k}=\dsum\limits_{(g,l)}J_{gl}e^{-i\mathbf{k}.(\mathbf{r}_{g}-\mathbf{r}%
_{l})}.  \label{11}
\end{equation}%
\ 

Using the low-$\mathbf{k}$ expression in two dimensions (as in Ref. 23) for $%
p=3$ (the ferromagnetic interaction is assumed to decay with the distance
between spins, $r$, as $J(r)=J/r^{p}$) we obtain: 
\begin{equation}
E_{k}=2mJ_{d}k(\pi ^{2}/8-\delta k).  \label{12}
\end{equation}

Using Eqs. (4), (8) and (9), we obtain the correlation function $%
\left\langle S^{-}S^{+}\right\rangle $; for spin $S=1/2$ we have $%
\left\langle S^{-}S^{+}\right\rangle =1/2-m$ and, therefore, the
magnetization is written in the form: 
\begin{equation}
m=\frac{1}{2}\left[ \frac{1}{N}\dsum\limits_{k}\coth (\beta E_{k}/2)\right]
^{-1}.  \label{13}
\end{equation}

\section{III. Results and discussion}

In the limit $m\rightarrow 0$, we find, from Eqs. (12) and (13), the
critical temperature ($T_{c}$), which is given by:%
\begin{equation}
\frac{k_{B}T_{c}}{J_{d}}=\left[ \frac{2}{N}\dsum\limits_{k}\frac{1}{k(\pi
^{2}/8-\delta k)}\right] ^{-1}.  \label{14}
\end{equation}%
In the thermodynamic limit ($N\rightarrow \infty $) one must replace the sum 
$\frac{1}{N}\dsum\limits_{k}\psi (\mathbf{k})$ by an integral $\frac{1}{%
(2\pi )^{d}}\dint\limits_{\text{1BZ}}\psi (\mathbf{k})d^{d}\mathbf{k}$ in
the $d$-dimensional $\mathbf{k}$-space, where 1BZ denotes the first
Brillouin zone. Then, the integral (14) is obtained by using the expansion
(12), and the critical temperature is given by%
\begin{equation}
\frac{k_{B}T_{c}}{J_{d}}=\frac{4\pi \delta }{\ln \left( \frac{\delta _{c}}{%
\delta _{c}-\delta }\right) }.  \label{15}
\end{equation}%
Note that the critical temperature vanishes at $\delta _{c}$, which, within
the present approximation, is given by $\delta _{c}=\pi /8\simeq 0.39$.

\FRAME{ftbpFU}{3.6305in}{2.7233in}{0pt}{\Qcb{Dependence of the reduced
critical temperature $k_{B}T_{c}/J_{d}$ on the parameter $\protect\delta $
for the quantum spin-$1/2$ Heisenberg model with ferromagnetic and dipolar
interactions with $p=3$. The insert is the behavior of $\protect\delta %
_{c}(p)$ as a function of the parameter $p$}}{}{Figure}{\special{language
"Scientific Word";type "GRAPHIC";maintain-aspect-ratio TRUE;display
"USEDEF";valid_file "T";width 3.6305in;height 2.7233in;depth
0pt;original-width 4.2376in;original-height 3.1704in;cropleft "0";croptop
"1";cropright "1";cropbottom "0";tempfilename
'IIWPAM01.wmf';tempfile-properties "XPR";}}

Numerical results for $T_{c}$ are shown in Fig. 1. The parameter $\delta $
is a measure of the strengthen of the frustration; therefore, long-range
order decreases as $\delta $ increases. Consequently, $T_{c}(\delta )$ goes
to zero at the critical value $\delta =\delta _{c}=\pi /8$. When the
interactions have the general form $J_{d}/r^{p}$, the existence of a
finite-temperature phase transition occurs for $2<p<4$ in two dimensions. By
numerically performing the sum in Eq. (13) for $2<p<4$ (see Ref. 23
for the general dispersion relation), we obtain the critical parameter $%
\delta _{c}(p)$ as a function of $p$ at $T=0$ (see the insert of Figure 1),
where above the curve the ordered phase is antiferromagnetic, while below it
the long-range order is ferromagnetic. We note that $\delta _{c}(p)$
increases as $p$ decreases, and when $p$ approaches $p=2$ we have a
divergence ($T_{c}$ also diverge) in $\delta _{c}(p=2)$ and around $p\simeq
3.6$ a minimum point appears. The point $p=4$ corresponds the to
Kosterlitz-Thouless phase transition \cite{23}, and with the presence of the
AF nearest-neighbor interaction the critical temperature vanishes at $\delta
_{c}(p=4)\simeq 0.793$. For $p>4$, the critical frustration parameter, $%
\delta _{c}(p)$, is zero, as physically expected.

\section{iV. Conclusions}

We study the phase diagram of the quantum spin-$1/2$ Heisenberg model with
competing interactions (i.e., presence of AF exchange and dipolar
interactions). The influence of the frustration is analyzed through the
variation of the parameter $\delta =J/J_{d}$. We observe that $T_{c}(\delta )
$ is null when we reach the critical value $\delta _{c}=\pi /8$. We also
analyzed the case of the general dependence of the long-range interaction in
the form $J_{ij}=J_{d}/r_{ij}^{p}$, and verified that $T_{c}$ tends to zero
at the critical parameter $\delta _{c}(p)$. The dependence of $%
\delta _{c}(p)$ on $p$ ($2<p<4$) present a divergence at $p=2$
($p\leq 2$ being the non-extensive regime) and a minimum point around $p\simeq 3.6$,
with finite value at $p=4$ (Kosterlitz-Thouless transition). Although
quantitative results are not expected to be very precise, within the
approximation we applied, we believe that we obtain reliable qualitative
results for the whole range of the parameters of the Hamiltonian. In
particular, our results agree with previous ones, when available.\newline

\begin{center}
\textbf{ACKNOWLEDGMENTS}
\end{center}

We would like to thank CNPq, FAPESC/FUNCITEC, and FAPEAM (Brazilian agencies) for the
financial support.


\begin{thebibliography}{99}
\bibitem{1} K. De Bell, A. B. MacIsaac, and J. P. Whitehead, \emph{Rev. Mod.
Phys}. \textbf{72}, 225 (2000), and references therein.

\bibitem{2} A. Scholl, J. St\"{o}hr, J. L\"{u}ning, J. W. Seo, J.
Fompeyrine, H. Siegwart, J. -P. Loequt, F. Nolting, S. Anders, E. E.
Fullerton, M. R. Scheinfein, and H. A. Padmore, \emph{Science} \textbf{287},
1014 (2000).

\bibitem{3} D. S. Deng, X. F. Jin, and Ruibao Tao, \emph{Phys. Rev. B} 
\textbf{65}, 132406 (2002).

\bibitem{4} N. D. Mermin and H. Wagner, \emph{Phys. Rev. Lett}. \textbf{17},
1133 (1966).

\bibitem{5} S. V. Maleev, \emph{Sov. Phys. JETP} \textbf{43}, 1240 (1976).

\bibitem{6} A. B. MacIsaac, J. P. Whitehead, M. C. Robinson, and K. De Bell, 
\emph{Phys. Rev. B} \textbf{51}, 16033 (1995).

\bibitem{7} A. Kashuba and V. L. Pokrovsky, \emph{Phys. Rev. Lett}. \textbf{%
70}, 3155 (1993).

\bibitem{8} P. M. Gleiser, F. A. Tamarit, and S. A. Cannas, \emph{Physica D} 
\textbf{168-169}, 73 (2002).

\bibitem{9} A. Mogni and G. Vertey, \emph{Phys. Rev. B} \textbf{61}, 3203
(2000).

\bibitem{10} L. C. Sampaio, M. P. de Albuquerque, and F. S. de Menezes, 
\emph{Phys. Rev. B} \textbf{54}, 6465 (1996).

\bibitem{11} J. H. Toloza, F. A. Tamarit, and S. A. Cannas, \emph{Phys. Rev.
B} \textbf{58}, R8885 (1998).

\bibitem{12} D. A. Stariolo and S. A. Cannas, \emph{Phys. Rev. B} \textbf{60}%
, 3013 (1999).

\bibitem{13} P. M. Gleiser, F. A. Tamarit, S. A. Cannas, and M. A.
Montemeiro, \emph{Phys. Rev. B} \textbf{68}, 1344011 (2003).

\bibitem{14} A. M. Abu-Labdeh, J. P. Whitehead, K. De Bell, and A. B.
MacIsaac, \emph{Phys. Rev. B} \textbf{65}, 244341 (2001).

\bibitem{15} D. S. Deng, X. F. Jin, and R. Tao, \emph{Phys. Rev. B} \textbf{%
69}, 1724031 (2004).

\bibitem{16} A. Aharony, et al., \emph{Phys. Rev. Lett}. \textbf{60}, 1330
(1988); K. B. Lyons, et al., \emph{Phys. Rev. Lett}. \textbf{60}, 732
(1988); J. M. Tranquada, et. al., \emph{Phys. Rev. Lett}. \textbf{60}, 2781
(1988).

\bibitem{17} Y. Endoh, et al., \emph{Phys. Rev. B} \textbf{37}, 7443 (1988);
A. P. Kampf, \emph{Phys. Rep}. \textbf{249}, 219 (1994); B. Keimer, et al.. 
\emph{Phys. Rev. B} \textbf{46}, 14034 (1992).

\bibitem{18} P. Chandra and B. Doucot, \emph{Phys. Rev. B} \textbf{38}, 9335
(1988).

\bibitem{19} T. W. Clinton, J. W. Lynn, J. Z. Liu, Y. X. Jia, and R. N.
Shelton, \emph{J. Appl. Phys}. \textbf{70}, 5751 (1991); D. M. Paul, H. A.
Mook, A. W. Hewat, B. C. Sales, L. A. Boatner, J. R. Thompson, and M.
Mosteller, \emph{Phys. Rev. B} \textbf{37}, 2341 (1988); H. A. Mook, D. M.
Paul, B. C. Sales, L. A. Batner, and L. Cussen, \emph{Phys. Rev. B} \textbf{%
38}, 12008 (1988); J. W. Lynn, T. W. Clinton, W. -H. Li, R. W. Erwin, J. Z.
Liu, K. Vandervoot, and R. N. Shelton, \emph{Phys. Rev. Lett}. \textbf{63},
2606 (1989).

\bibitem{20} D. N. Zubarev, \emph{Usp. Fiz. Nauk}. \textbf{71}, 71 (1960);
see also S. V. Tayablikov, \emph{Methods in the Quantum Theory of Magnetism}
(Plenum, New York, 1967).

\bibitem{21} M. Hamedoun, Y. Cherriet, A. Hourmatallah, and N. Benzokour, 
\emph{Phys. Rev. B} \textbf{63}, 1724021 (2001).

\bibitem{22} A. Cavallo, F. Cosenza, and L. De Cesare, \emph{Phys. Rev. B} 
\textbf{66}, 1744391 (2002).

\bibitem{23} H. Nakano and M. Takahashi, \emph{Phys. Rev. B} \textbf{52},
6606 (1995).

\bibitem{24} P. Bruno, \emph{Phys. Rev. Lett.} \textbf{87}, 137203 (2001).

\bibitem{25} O. N. Vassiliev, I. V. Rojdestvenski, and M. G. Cottam, \emph{%
Physica A} \textbf{294}, 139 (2001).
\end{thebibliography}
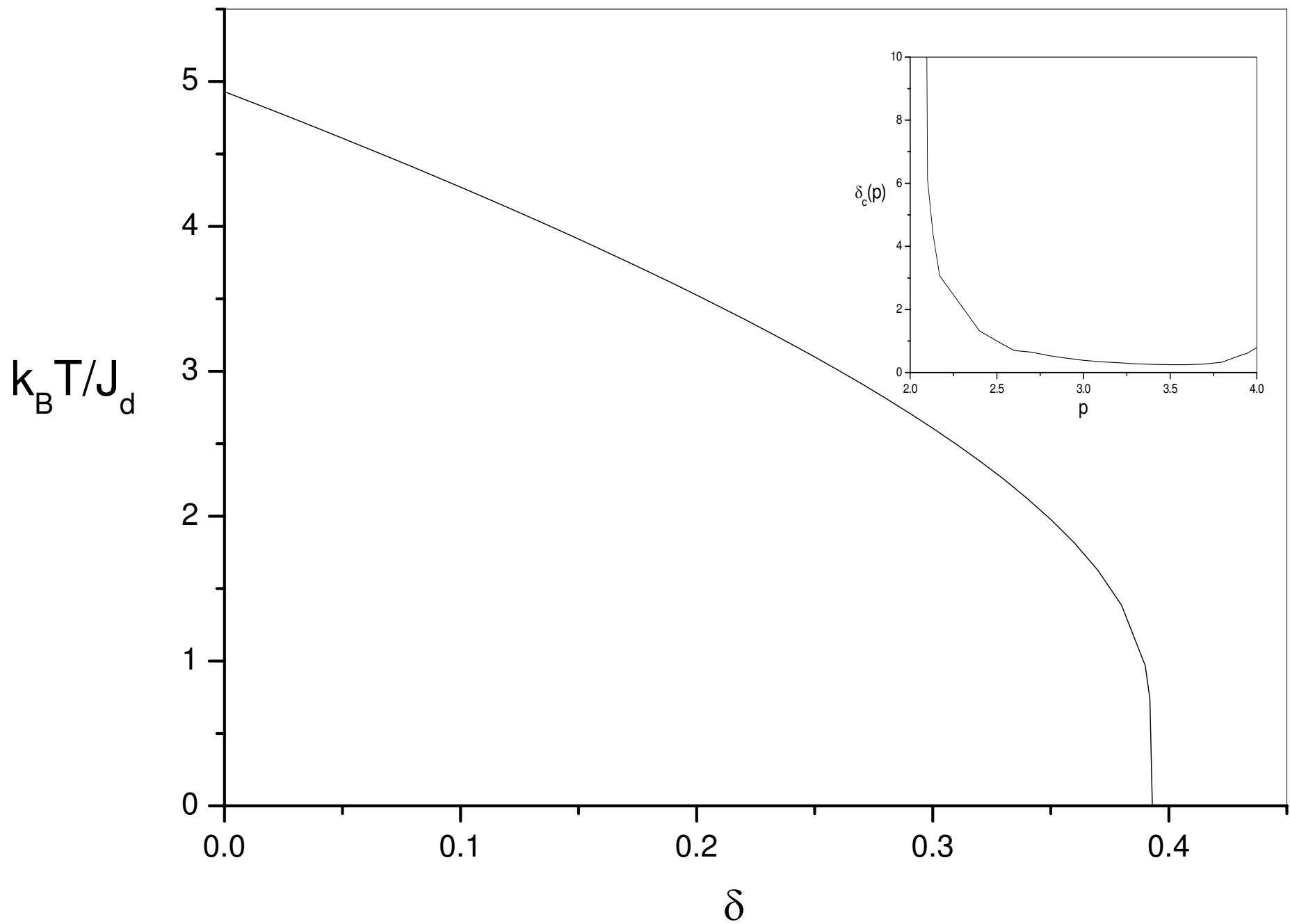
\end{document}